\documentclass[epj]{svjour}
\usepackage{graphics}
\begin{document}
\title{Phase structure of intrinsic curvature models on dynamically triangulated disk with fixed boundary length}
\author{Hiroshi Koibuchi% etc
%\thanks{\emph{Present address:Nakane 866, Hitachinaka, Ibaraki 312-8508, Japan } }%
}                     % Do not remove
%
%\offprints{koibuchi@mech.ibaraki-ct.ac.jp}          % Insert a name or remove this line
%
\institute{Department of Mechanical and Systems Engineering \\
  Ibaraki National College of Technology \\
  Nakane 866, Hitachinaka, Ibaraki 312-8508, Japan }
%
%\date{Received:  Revised version:  }
\abstract{
A first-order phase transition is found in two types of intrinsic curvature models defined on dynamically triangulated surfaces of disk topology. The intrinsic curvature energy is included in the Hamiltonian. The smooth phase is separated from a non-smooth phase by the transition. The crumpled phase, which is different from the non-smooth phase, also appears at sufficiently small curvature coefficient $\alpha$. The phase structure of the model on the disk is identical to that of the spherical surface model, which was investigated by us and reported previously. Thus, we found that the phase structure of the fluid surface model with intrinsic curvature is independent of whether the surface is closed or open.}
 %end of abstract
\authorrunning {H.Koibuchi}
\titlerunning {Phase structure of fluid membrane models with intrinsic curvature}
\maketitle
%
%----------------------------------------------------------
\section{Introduction}\label{intro}
%----------------------------------------------------------
Triangulated surfaces are one of the basic models to investigate the physics for biological membranes and strings \cite{NELSON-SMMS-2004,Gompper-Schick-PTC-1994,BOWICK-TRAVESSET-PREP-2001,WIESE-PTCP19-2000,WHEATER-JP-1994}.  Theoretical investigations, on the basis of renormalization group strategy, have been made to reveal how the fluctuations of surfaces stiffen/soften the surface \cite{DavidGuitter-EPL-1988,Peliti-Leibler-PRL-1985,BKS-PLA-2000,BK-PRB-2001,Kleinert-EPJB-1999}.  A considerable number of numerical studies have also been performed for understanding the phase structure by using the surface simulations on fluid (or dynamical connectivity) surfaces \cite{CATTERALL-NPB-SUPL-1991,AMBJORN-NPB-1993,BCHGM-NPB-1993,ABGFHHE-PLB-1993,KOIB-PLA-2002,KOIB-PLA-2004,KOIB-EPJB-2005,KOIB-PLA-2003-2,KOIB-EPJB-2004} and on tethered (or fixed connectivity) surfaces \cite{KOIB-EPJB-2004,KOIB-PLA-2005-1,KOIB-PLA-2006-1,BCFTA-1996-1997,WHEATER-NPB-1996,KANTOR-NELSON-PRA-1987,GOMPPER-KROLL-PRE-1995,KOIB-PRE-2003,KOIB-PRE-2004-1,Kownacki-Diep-PRE-2002,KOIB-PRE-2004-2,KOIB-PRE-2005,KOIB-NPB-2006,KANTOR-SMMS-2004,KANTOR-KARDER-NELSON-PRA-1987,BCTT-EPJE-2001,BOWICK-SMMS-2004,NELSON-SMMS-2004-2}. The fluid model is defined on dynamically triangulated surfaces in the framework of the surface simulations, whereas the tethered model is defined on fixed connectivity surfaces. The surface simulations have relatively long history rather than that of another numerical technique Molecular Dynamics simulations for the lipid models for biological membranes \cite{Goetz-Lipowsky-JCP-1998,Noguchi-Takatsu-JCP-2001,Noguchi-Takatsu-BJ-2002,Deserno-PRE-2005}. The extrinsic curvature energy is always included in the surface Hamiltonian, because it is  considered to smoother the surface. However, we must recall that the intrinsic curvature energy such as the deficit angle term can also be assumed \cite{BJ-PRD-1994,BEJ-PLB-1993,BIJJ-PLB-1994} in the Hamiltonian because it can smoother the surface. 

It has been shown recently by Monte Carlo (MC) simulations that the extrinsic curvature model of Helfrich and Polyakov-Kleinert \cite{HELFRICH-NF-1973,POLYAKOV-NPB-1986,Kleinert-PLB-1986} undergoes a discontinuous transition between the smooth phase and the crumpled phase on fixed connectivity spherical surfaces \cite{Kownacki-Diep-PRE-2002,KOIB-PRE-2004-2,KOIB-PRE-2005}. Moreover, the phase transition occurs independent of whether the Gaussian bond potential is included in the Hamiltonian or not \cite{KOIB-NPB-2006}. It was also shown that the transition occurs independent of how the extrinsic curvature term is discretized \cite{KOIB-NPB-2006}. It has also been reported that a tethered surface model undergoes a first-order transition, where the Hamiltonian includes an intrinsic curvature term \cite{KOIB-EPJB-2004,KOIB-PLA-2005-1,KOIB-PLA-2006-1}. The transition in this model occurs independent of whether the surface is closed or open. In fact, the transition can be observed on a sphere \cite{KOIB-EPJB-2004}, on a disk \cite{KOIB-PLA-2005-1}, and on a torus \cite{KOIB-PLA-2006-1}. 

However, it seems that many points remain to be studied in the surface models. We consider that one interesting problem is how the fluidity of lateral diffusion influences the phase transition in the model with some curvature Hamiltonian \cite{CATTERALL-NPB-SUPL-1991,AMBJORN-NPB-1993,BCHGM-NPB-1993,ABGFHHE-PLB-1993,KOIB-PLA-2002,KOIB-PLA-2004,KOIB-EPJB-2005,KOIB-PLA-2003-2,KOIB-EPJB-2004}. The phase diagram of the intrinsic curvature model was studied on dynamically triangulated spheres \cite{KOIB-EPJB-2004}, however, it has not been studied yet on dynamically triangulated surfaces of disk topology. 

If a membrane model undergoes a phase transition on dynamically triangulated surfaces, we must have reasons why such transition occurs. In fact, it is widely accepted that we can see no phase transition in the fluid membranes. The standard argument for fluid membranes is that there is no long-range order on the surface at finite bending rigidity \cite{DeGennes-Taupin-JPC-1982}. 

First of all, we must not forget that the transition seems to occur on relatively small surfaces \cite{CATTERALL-NPB-SUPL-1991,AMBJORN-NPB-1993,BCHGM-NPB-1993,ABGFHHE-PLB-1993}. Even if the model in this paper undergoes a transition on small surfaces, the transition may disappear on sufficiently large surfaces. Secondly, it is possible that the transition occurs without the smooth phase: the transition is not the one between the smooth phase and the crumpled phase but the one between the crumpled phase and the folded phase, which are respectively characterized by the Hausdorff dimension $2\!<\!H\!<\!3$ and $H\!\geq\!3$ in the case of self-intersecting surfaces. We have some other possibilities of the origin including the argument of Helfrich that the thermal fluctuation increases the effective bending rigidity \cite{Pinnow-Helfrich-EPJE-2000}. It is possible that the curvature coefficient is increased by the fluctuation of surfaces just like the case for the coefficient of the Gaussian curvature \cite{BOWICK-SMMS-2004}, which is an intrinsic curvature as well as the deficit angle term in this paper. 

In this paper two types of fluid surface models are studied by MC simulations on a disk with intrinsic curvature. The Hamiltonian of the first model is a linear combination of the Gaussian bond potential and an intrinsic curvature term. The second model is a tensionless model, whose Hamiltonian is given by a linear combination of the intrinsic curvature term and a hard-wall potential, which gives only an upper bound on the bond length.

The simulations are performed under fixed total number of vertices on the boundary of disk as well as on the interior of disk. As a consequence, the mean boundary length is fixed to be $n_{\rm B} \langle \ell\rangle$, where $n_{\rm B}$ is the total number of bonds on the boundary, and $\langle \ell\rangle$ is the mean bond length.

It will be shown that both models undergo a first-order transition between the smooth phase and a non-smooth phase on relatively large surfaces. Thus, we can exclude the above-mentioned first and the second possibilities of the reason for the transition in the fluid membrane model with intrinsic curvature. As a consequence, the result in this paper suggests that the fluctuation of surfaces does not always soften the surface. Thus we understand that fluidity of dynamically triangulated surfaces does not eliminate the long-range order whenever the surface model is defined with intrinsic curvature. This is in sharp contrast to the case in the model with extrinsic curvature, where the surface is softened by the surface fluctuations. 

Combining the results in this paper and those in Ref. \cite{KOIB-EPJB-2004}, we will confirm that the transition is independent of whether the surfaces is closed  or open in fluid and intrinsic-curvature models. We consider that this result is remarkable because phase transitions are always severely influenced by boundary conditions that restrict the value of variables in statistical models. 
%------------------------------------------
\section{Models}\label{model}
%------------------------------------------
%++++++++++++++++++++++++++++++++++
\begin{figure}[hbt]
\centering
\resizebox{0.25\textwidth}{!}{%
  \includegraphics{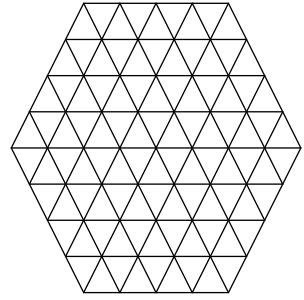}
}
\caption{A triangulated disk of size $(N,N_E,N_P)\!=\!(61,156,96)$, where $N,N_E,N_P$ are the total number of vertices, the total number of edges, the total number of plaquettes (or triangles).}
\label{fig-1}
\end{figure}
%++++++++++++++++++++++++++++++++++
The starting configuration for MC simulations is a flat and regularly triangulated disk, which is obtained by dividing the hexagon. Figure \ref{fig-1} is a lattice of size $N\!=\!61$, which is obtained by dividing every edge of the original hexagon into $4$-pieces. By dividing the edges into $L$-pieces we have a lattice of $(N,N_E,N_P)\!=\!(3L^2\!+\!3L\!+\!1, 9L^2\!+\!3L,6L^2)$, where $N,N_E,N_P$ are the total number of vertices, the total number of edges, and the total number of plaquettes (or triangles).

The Gaussian bond potential $S_1$, the intrinsic curvature $S_3$, and the hard-wall potential $V$ are given by 
\begin{eqnarray}
\label{S1S3V}
&&S_1=\sum_{(ij)} \left(X_i-X_j\right)^2, \quad S_3= \sum_i \left( \delta_i - \delta_0 \right)^2, \nonumber \\
&&V=\sum_{(ij)} V(|X_i-X_j|),  
\end{eqnarray}
where $\sum_{(ij)}$ is the sum over all bonds $(ij)$ connecting the vertices $X_i$ and $X_j$, and  $\delta_i$ in $S_3$ is the sum of angles of the triangles meeting at the vertex $i$. We note that $\sum_{(ij)}$ in $S_1$ and $V$ include the sum over the boundary bonds. As a consequence, the length of the boundary bonds can vary in the simulations. The value of the symbol $\delta_0$ in $S_3$ is given by 
\begin{equation}
\label{delta-0}
\delta_0=\left\{ \begin{array}{ll}
    2\pi                   & \mbox{(internal vertices);} \\
    \pi (\mbox{or}\; 2\pi/3) & \mbox{(boundary vertices),} 
   \end{array} \right.
\end{equation}
where $2\pi/3$ is assigned to the six vertices, and $\pi$ to the remaining vertices on the boundary. Because of the definition of $\delta_0$ in (\ref{delta-0}), we can call $S_3$ in Eq.(\ref{S1S3V}) as a deficit angle term. 

The symbol $V(|X_i\!-\!X_j|)$ in the hard-wall potential $V$ in Eq. (\ref{S1S3V}) is the potential between the vertices $i$ and $j$, and is defined by
\begin{equation}
\label{V}
V(|X_i-X_j|)= \left\{
       \begin{array}{@{\,}ll}
       0 & \quad (|X_i-X_j| < r_0), \\
      \infty & \quad ({\rm otherwise}). 
       \end{array}
       \right. 
\end{equation}
The value of $r_0$ in the right hand side of Eq. (\ref{V}) is fixed to $r_0^2\!=\!1.08$. Then we have $\langle \sum (X_i\!-\!X_j)^2 \rangle /N \simeq 3/2$, which is satisfied when the Gaussian bond potential $S_1\!=\!\sum (X_i\!-\!X_j)^2$ is included in the Hamiltonian, which includes no hard-wall potential such as $V$ in Eq.(\ref{S1S3V}). 

The constraint $|X_i\!-\!X_j| \!<\! r_0$ is necessary for the bond length to have a well-defined value if the Gaussian bond potential $S_1$ is not included in the Hamiltonian. In fact, the size of the surface grows larger and larger in the MC simulations if it were not for the constraint $|X_i\!-\!X_j| \!<\! r_0$ in the hard-wall potential. As a consequence, the model seems to depend on a hidden length scale introduced by $r_0$, and then the model appears to be ill-defined. However, we checked in \cite{KOIB-PLA-2003-2} that there is no $r_0$-dependence on the results. This is a consequence of the scale invariant property of the model. Therefore, we use $r_0^2\!=\!1.08$ in the MC simulations.

The partition function is defined by 
\begin{equation}
 \label{partition-function}
Z(\alpha) = \sum_T \int \prod _{i=1}^N dX_i \exp\left[-S(X,T)\right],
\end{equation}
where $N$ is the total number of vertices as described above, and $\sum_T$ denotes the summation over all possible triangulations $T$ under a constraint which will be described in the next section. The expression $S(X,T)$  shows that $S$ explicitly depends on the variable $X$ and $T$. The coefficient $\alpha$ is the elastic modulus. The surfaces are allowed to self-intersect. The center of surface is fixed to remove the translational zero-mode. 

Two types of models, which we call {\it model 1} and {\it model 2}, are defined by Hamiltonians
\begin{eqnarray}
 \label{Hamiltonians}
&&S(X,T)=S_1 + \alpha S_3  \qquad({\rm model \;1}),\\
&&S(X,T)=V+ \alpha S_3  \qquad({\rm model\; 2}).
\end{eqnarray}

 It should be noted that the Gaussian bond potential $S_1$ per vertex at the internal vertices differs from the one at the boundary vertices. In fact, the co-ordination number is $q\!=\!6$ at the internal vertices and $q\!=\!3$ or $q\!=\!4$ at the boundary vertices. The intrinsic curvature $S_3$ at the internal vertices can also be different from those at the boundary ones due to the same reason for $S_1$. This difference of $S_1$ per vertex (or $S_3$ per vertex)  is typical to the model on the open disk, and such large difference cannot be seen in the model on closed surfaces such as a sphere or a torus. It must also be noted that the boundary is not completely free but shares both of the Gaussian bond potential (or the hard-wall potential) and the intrinsic curvature energy in the models of this paper.

We comment on a relation of $S_3$ in Eq.(\ref{S1S3V}) to an intrinsic curvature energy $S_3\!=\!-\sum_i \log(\delta_i/2\pi)$ in \cite{KOIB-EPJB-2004}. This $S_3$ in  \cite {KOIB-EPJB-2004} is closely related with the co-ordination dependent term  $S_3\!=\!-\sum_i \log (q_i/6)$, which comes from the integration measure $\prod_i dX_i q_i^\alpha$ \cite{DAVID-NPB-1985} in the partition function for the model on a sphere. $S_3\!=\!-\sum_i \log(\delta_i/2\pi)$ has the  minimum value when $\delta_i\!=\!2\pi$ is satisfied for all $i$, and hence it becomes smaller on a smooth sphere than on a crumpled one. However, the flat configuration of the hexagonal lattice shown in Fig.\ref{fig-1} does not minimize $S_3\!=\!-\sum_i \log(\delta_i/2\pi)$, because it includes vertices such that $\delta_i\!\not=\!2\pi$. This is the reason why we use $S_3$ in Eq.(\ref{S1S3V}), which becomes zero on such flat configuration as shown in Fig.\ref{fig-1}. 

%------------------------------------------
\section{Monte Carlo technique}\label{MC-Techniques}
%------------------------------------------
The canonical Monte Carlo technique is used to update the variables $X$ to a new $X^\prime$ so that $X^\prime \!=\! X \!+\! \delta X$, where the small change $\delta X$ is made at random in a small sphere in ${\bf R}^3$. The new position $X^\prime$ is accepted with the probability ${\rm Min} [1, \exp (-\Delta S)]$, where $ \Delta S \!=\!S({\rm new})\!-\!S({\rm old})$. The radius $\delta r$ of the small sphere is chosen at the beginning of the simulations to maintain the rate of acceptance $r_X$ for the $X$-update as $0.4 \leq r_X \leq 0.6$. 

The summation over $T$ in $Z$ of Eq.(\ref{partition-function}) is performed by bond flips. The bonds are labeled by sequential numbers. A bond is randomly chosen, and the flip is accepted with the probability ${\rm Min} [1, \exp(-\Delta S)]$. The bond flip is applied only to internal bonds except the bonds whose both end points touch the boundary after the flip. The boundary bonds remain unflipped.

$N$-updates for $X$ and $N$-updates for $T$ are consecutively performed, and these make one MCS (Monte Carlo Sweep). We use a random number called Mersenne Twister \cite{Matsumoto-Nishimura-1998} in the MC simulations. The above mentioned Metropolis procedure is applied only for $X$ and $T$, which satisfy the constraint $|X_i\!-\!X_j| \!<\! r_0$ in Eq.(\ref{V}) for model 2.   

We use surfaces of size $N\!=\!817$, $N\!=\!1657$, $N\!=\!3367$, and  $N\!=\!5167$, which correspond to  the partitions $L\!=\!16$, $L\!=\!23$, $L\!=\!33$, and  $L\!=\!41$ respectively.  The size of the surfaces is relatively larger than that of spherical surfaces used in \cite{KOIB-EPJB-2004}. 

The total number of MCS after the thermalization is about $1.0\!\times\!10^8\sim 3.0\!\times\!10^8$ in the smooth phase close to the transition point of the surfaces of size $N\!=\!5167$,  $N\!=\!3367$, and   $N\!=\!1657$,  and $1.6\!\times\!10^8$ for  $N\!=\!817$. Relatively smaller number of MCS ($1.0\!\times\!10^8\sim 1.6\!\times\!10^8$) (including the thermalization MCS) is done at non-transition point. The thermalization MCS is  $0.5\!\times\!10^7\sim 1.0\!\times\!10^8$, which depends on $N$ and $\alpha$. Very large number of thermalization MCS is sometimes needed close to the transition point, because there are smooth surfaces that start to collapse into non-smooth surfaces after a long MC run.  Physical quantities are calculated at every 500 MCS after the thermalization.

%------------------------------------------
\section{Results}\label{Results}
%------------------------------------------
%++++++++++++++++++++++++++++++++++
\begin{figure}[hbt]
%\vspace{1cm}
%WinTpicVersion3.08
\unitlength 0.1in
\begin{picture}( 0,0)(  20,30)
\put(8,27){\makebox(0,0){(a) Model 1 ($\alpha\!=\!2$)}}%
\put(25,27){\makebox(0,0){(b) Model 1 ($\alpha\!=\!22$)}}%
\put(28,-10){\makebox(0,0){(d) Sections of the surface in (c)}}%
\put(8,-10){\makebox(0,0){(c) Model 1 ($\alpha\!=\!28$)}}%
\end{picture}%
\vspace{1cm}
\centering
\resizebox{0.49\textwidth}{!}{%
  \includegraphics{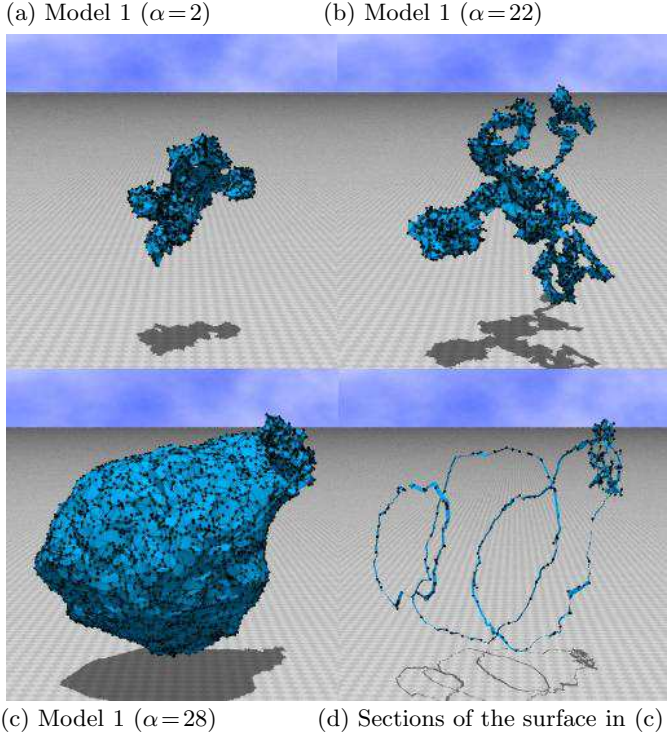}
}
\vspace{0.5cm}
\caption{Snapshots of model 1 surfaces of size $N\!=\!5167$ at (a) $\alpha\!=\!2$ (crumpled), (b) $\alpha\!=\!22$ (non-smooth), and (c) $\alpha\!=\!28$ (smooth), and (d) sections of the surface in (c). Figures are shown in the same scale. }
\label{fig-2}
\end{figure}
%++++++++++++++++++++++++++++++++++
Snapshots of model 1 surfaces are shown in Figs.\ref{fig-2}(a), \ref{fig-2}(b), and \ref{fig-2}(c), which are obtained at $\alpha\!=\!2$, $\alpha\!=\!22$, and $\alpha\!=\!28$, respectively. Figure \ref{fig-2}(d) is a set of sections of the surface in Fig.\ref{fig-2}(c). The figures are shown in the same scale, and the size is $N\!=\!5167$. It is easy to see that the surface in Fig.\ref{fig-2}(a) is crumpled (or folded), and the surface in Fig.\ref{fig-2}(b) is a branched polymer like one. The surface in Fig.\ref{fig-2}(c) is obviously smooth, and it looks like a sphere with a hole. The surface in Fig.\ref{fig-2}(c) is separated from that in Fig.\ref{fig-2}(b) by a first-order transition as we will see below.

%++++++++++++++++++++++++++++++++++
\begin{figure}[hbt]
%\vspace{1cm}
%WinTpicVersion3.08
\unitlength 0.1in
\begin{picture}( 0,0)(  20,30)
\put(8,27){\makebox(0,0){(a) Model 2 ($\alpha\!=\!10$)}}%
\put(25,27){\makebox(0,0){(b) Model 2 ($\alpha\!=\!32$)}}%
\put(25,-10.5){\makebox(0,0){(d) Model 2 ($\alpha\!=\!44$)}}%
\put(8,-10.5){\makebox(0,0){(c) Model 2 ($\alpha\!=\!36$)}}%
\end{picture}%
\vspace{1cm}
\centering
\resizebox{0.49\textwidth}{!}{%
  \includegraphics{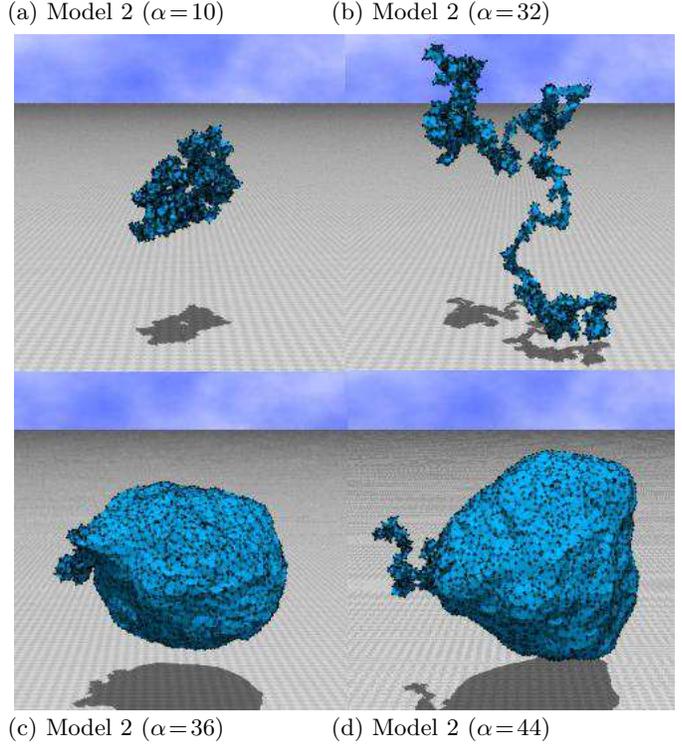}
}
\vspace{0.5cm}
\caption{Snapshots of model 2 surfaces of size $N\!=\!5167$ at (a) $\alpha\!=\!10$ (crumpled), (b) $\alpha\!=\!32$ (non-smooth), (c) $\alpha\!=\!36$ (transient surface between non-smooth and smooth), and (d) $\alpha\!=\!44$ (smooth). The surface in (c) is almost smooth. Figures are shown in the same scale.}
\label{fig-3}
\end{figure}
%++++++++++++++++++++++++++++++++++
Snapshots of model 2 surfaces are shown in Figs.\ref{fig-3}(a), \ref{fig-3}(b), \ref{fig-3}(c), and \ref{fig-3}(d), which are obtained at $\alpha\!=\!10$, $\alpha\!=\!32$, $\alpha\!=\!36$, and $\alpha\!=\!44$, respectively. The figures are shown in the same scale, and the size is $N\!=\!5167$.  It is also easy to see that the surface in Fig.\ref{fig-3}(a) is almost folded, and that the surface in  Fig.\ref{fig-3}(b) is a branched polymer like one, which is rather oblong than the one in Fig.\ref{fig-2}(b) of model 1.  The snapshot in Fig.\ref{fig-3}(c) shows an intermediate and transient state between the non-smooth phase and the smooth phase. The reason why the surface in Fig.\ref{fig-3}(c) is considered to be an intermediate one is because the bending energy $S_2$ is relatively large, which will be shown in the next paragraph. We can see from Fig.\ref{fig-3}(c) that non-smooth part is almost shrunk. It is expected that the surface becomes a smooth surface after a long MC run. Because the smooth phase is separated from the non-smooth phase by first-order transition, we expect that the configuration in Fig.\ref{fig-3}(c) eventually belongs to the smooth phase. On the contrary, the surface in Fig.\ref{fig-3}(d) is almost smooth, and the non-smooth part seems to be shrunk. We note that the surface seems to be closed in the smooth phase as that in Fig.\ref{fig-2}(c) of model 1. The surface in Fig.\ref{fig-3}(d) is separated from that in Fig.\ref{fig-3}(b) by a first-order transition as we will see below.

From the bending energy $S_2$ we can extract information about how smooth the surface is, although $S_2$ is not included in the Hamiltonian. The bending energy $S_2$ is defined by
\begin{equation}
\label{S2}
S_2=\sum_{ij}\left(1-{\bf n}_i\cdot{\bf n}_j \right),
\end{equation}
where ${\bf n}_i$ is the unit normal vector of the triangle $i$. The symbol $ij$ in $\sum_{ij}$ denotes the nearest-neighbor triangles $i$ and $j$, which have a common bond. The pair ${\bf n}_i\cdot{\bf n}_j$ is not defined on the boundary bonds, because we have no triangle outside the boundary. 

%++++++++++++++++++++++++++++++++++
\begin{figure}[hbt]
\centering
\resizebox{0.49\textwidth}{!}{%
  \includegraphics{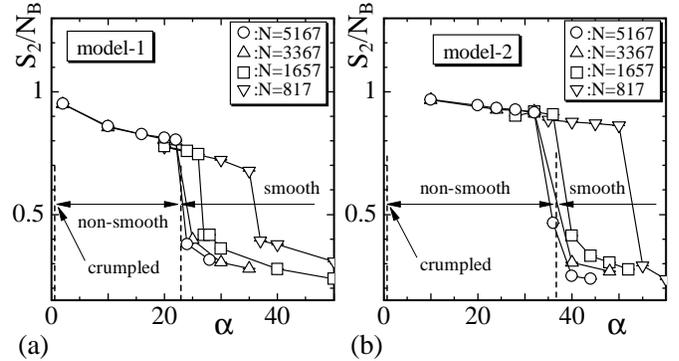}
}
\caption{$S_2/N_B$ vs. $\alpha$ obtained in (a) model 1 and (b) model 2. $N_B$ is the total number of internal bonds. The size of surfaces are $N\!=\!817$, $N\!=\!1657$, $N\!=\!3397$, are $N\!=\!5167$. Dashed lines denote the phase boundaries of the models on the $N\!=\!5167$ surface. }
\label{fig-4}
\end{figure}
%++++++++++++++++++++++++++++++++++
Figures \ref{fig-4}(a) and \ref{fig-4}(b) are plots of the bending energy $S_2/N_B$ against $\alpha$ obtained in model 1 and model 2, respectively. $N_B$ is the total number of internal bonds. We find from Figs.\ref{fig-4}(a) and \ref{fig-4}(b) that $S_2/N_B$ is obviously discontinuous at finite $\alpha$. Thus, we have shown that both models undergo a phase transition between the smooth phase and the non-smooth phase. As we have seen in the snapshots, both model have three different phases; the smooth, the non-smooth, and the crumpled, which is expected to emerge at sufficiently small $\alpha$. 

One dashed line in the figures denotes the phase boundary between the smooth phase and the non-smooth phase on the $N\!=\!5167$ surface, and the other one denotes the phase boundary between the non-smooth phase and the crumpled phase. Since the transition point of the $N\!=\!3367$ surface is almost identical to that of the $N\!=\!5167$ surface, the transition seems not to be influenced by the finite-size effects in those system size. We note also that the non-smooth surfaces do not appear between the smooth phase and the crumpled phase on the $N\!=\!817$ and $N\!=\!1657$ surfaces. The reason of this seems due to the finite-size effect. 

The order of the transitions between the smooth phase and the non-smooth one is considered as first order, because $S_2/N_B$ is discontinuous as shown above. We recall that a phase transition is called a first-order (or discontinuous) one if some physical quantity discontinuously changes even when the quantity is not included in the Hamiltonian. On the contrary, we can see no transition between the crumpled and the non-smooth phases. In fact, $S_2/N_B$ and some other physical quantities smoothly vary at the transition region of $\alpha$, just as in the spherical surface model \cite{KOIB-EPJB-2004}.

%++++++++++++++++++++++++++++++++++
\begin{figure}[hbt]
\centering
\resizebox{0.49\textwidth}{!}{%
  \includegraphics{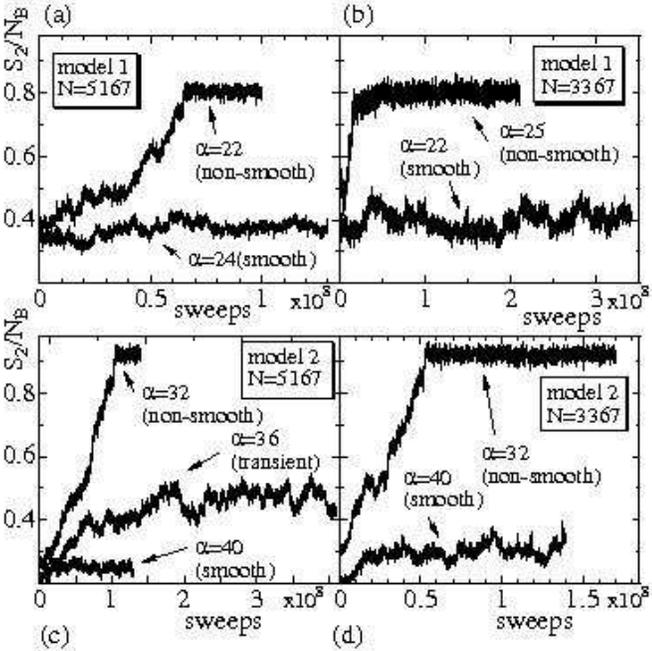}
}
\caption{Variation of $S_2/N_B$ against MCS obtained on model 1 surface of size (a) $N\!=\!5167$, (b) $N\!=\!3367$, and on model 2 surface of (c) $N\!=\!5167$, (d) $N\!=\!3367$.  }
\label{fig-5}
\end{figure}
%++++++++++++++++++++++++++++++++++
In order to see how is the convergence of the simulations close to the transition point, we plot the variation of $S_2/N_B$ against MCS including the thermalization MCS in Figs.\ref{fig-5}(a)--\ref{fig-5}(d): Fig.\ref{fig-5}(a) is $S_2/N_B$ obtained at $\alpha\!=\!22$(non-smooth) and $\alpha\!=\!24$(smooth) on the $N\!=\!5167$ surface of model 1, Fig.\ref{fig-5}(b) is those obtained at $\alpha\!=\!25$(non-smooth) and $\alpha\!=\!22$(smooth) obtained on the $N\!=\!3367$ surface of model 1, Fig.\ref{fig-5}(c) is those obtained at $\alpha\!=\!32$(non-smooth), $\alpha\!=\!36$(transient), and $\alpha\!=\!40$(smooth) on the $N\!=\!5167$ surface of model 2, and Fig. \ref{fig-5}(d) is those obtained at $\alpha\!=\!32$(non-smooth) and $\alpha\!=\!40$(smooth) obtained on the $N\!=\!3367$ surface of model 2. 
We immediately see that the total number of MCS after the thermalization in the non-smooth phase is relatively smaller than that in the smooth phase in the $N\!=\!5167$ surface shown in Figs.\ref{fig-5}(a) and \ref{fig-5}(c). The reason of this is because the value of $S_2/N_B$ is stable in the non-smooth phase after the thermalization. On the contrary, relatively large number of MCS was performed in the smooth phase close to the transition point, because a smooth surface can collapse into a non-smooth surface after a long MC run as stated in the final paragraph of the previous section. The variation at $\alpha\!=\!36$(transient) in Fig.\ref{fig-5}(c) indicates that the surface seems belong to the smooth phase after long MC run, although the value of $S_2/N_B$ still remains relatively larger than that of $\alpha\!=\!40$(smooth). 

We found no variation of $S_2/N_B$ against MCS indicating jumps from the smooth phase to the non-smooth phase and vice versa. However, the value of $S_2/N_B$ in the non-smooth phase is quite distinct from that in the smooth phase after the thermalization. Therefore, the gaps (or discontinuities) in $S_2/N_B$ shown in Figs.\ref{fig-4}(a) and \ref{fig-4}(b) are considered as a sign of a discontinuous transition.

%++++++++++++++++++++++++++++++++++
\begin{figure}[hbt]

\centering
\resizebox{0.49\textwidth}{!}{%
  \includegraphics{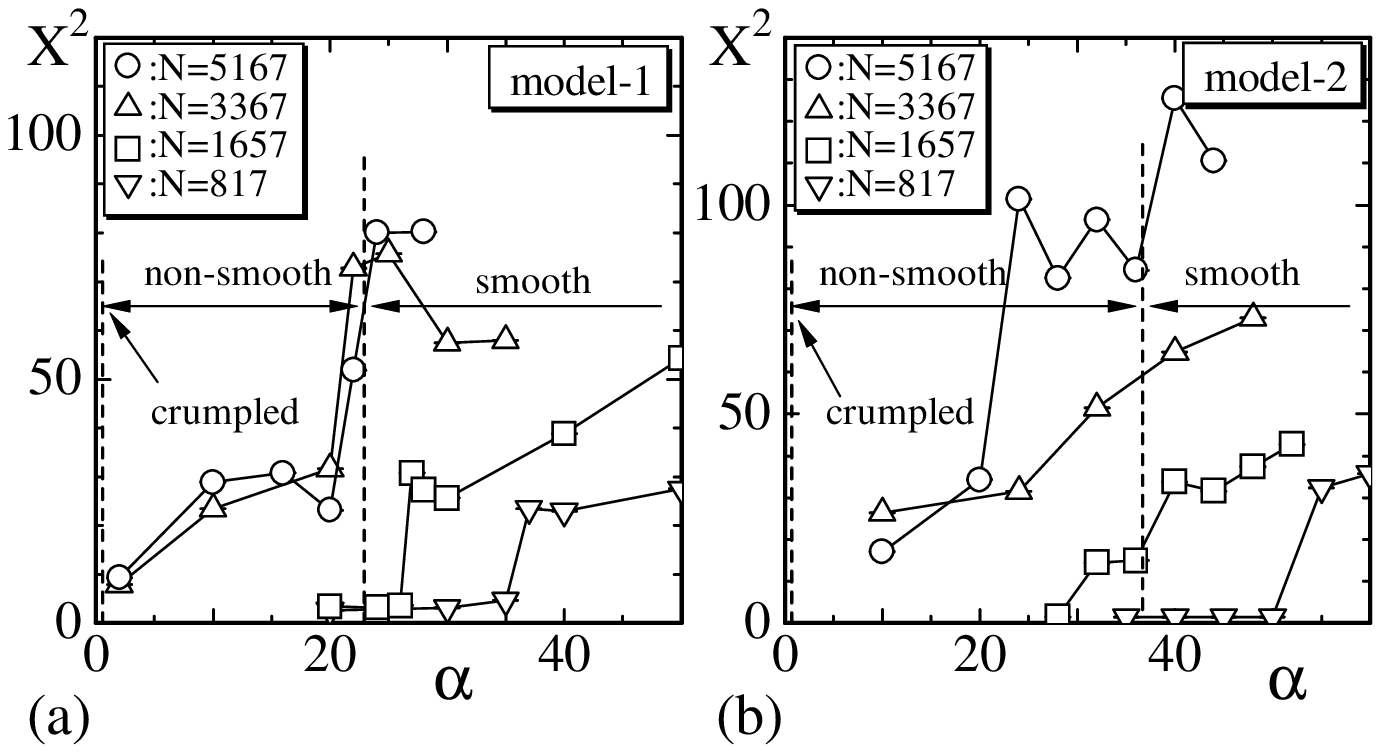}
}
\caption{$X^2$ vs. $\alpha$ obtained in (a) model 1 and (b) model 2. The size of surfaces  is  $N\!=\!817$, $N\!=\!1657$,  $N\!=\!3367$, and $N\!=\!5167$. Dashed lines denote the phase boundaries of the models on the $N\!=\!5167$ surface.}
\label{fig-6}
\end{figure}
%++++++++++++++++++++++++++++++++++
In order to see the size of surfaces, we calculate the mean square size $X^2$ defined by
\begin{equation}
\label{X2}
X^2={1\over N} \sum_i \left(X_i-\bar X\right)^2, \quad \bar X={1\over N} \sum_i X_i,
\end{equation}
where $\bar X$ is the center of the surface.

We plot $X^2$ against $\alpha$ in Figs.\ref{fig-6}(a) and \ref{fig-6}(b) obtained in model 1 and model 2, respectively. The dashed lines drawn vertically in the figures denote the phase boundaries of the $N\!=\!5167$ surface. These lines were drawn at the same position as the corresponding dashed lines in Figs.\ref{fig-4}(a) and \ref{fig-4}(b).  We see in Fig.\ref{fig-6}(a) that $X^2$ discontinuously changes at finite $\alpha$ in model 1. The transition point $\alpha$, where $X^2$ discontinuously changes in Fig.\ref{fig-6}(a), is identical to $\alpha$ where $S_2/N_B$ discontinuously changes in Fig.\ref{fig-4}(a). The transition point $\alpha$ shown in Fig.\ref{fig-6}(a) distinguishes the smooth phase from the non-smooth phase in model 1. This indicates that the surface in the non-smooth phase is almost crumpled.

On the contrary, these two phases can hardly be distinguished by $X^2$ in model 2, because $X^2$ in the non-smooth phase is almost comparable to $X^2$ in the smooth phase close to the transition point. This phenomenon is apparent on the $N\!\geq\!1657$ surfaces and consistent with the fact that the surface of the non-smooth phase in model 2 is relatively oblong and extended as shown in Fig.\ref{fig-3}(b). 

%++++++++++++++++++++++++++++++++++
\begin{figure}[hbt]
\centering
\resizebox{0.49\textwidth}{!}{%
  \includegraphics{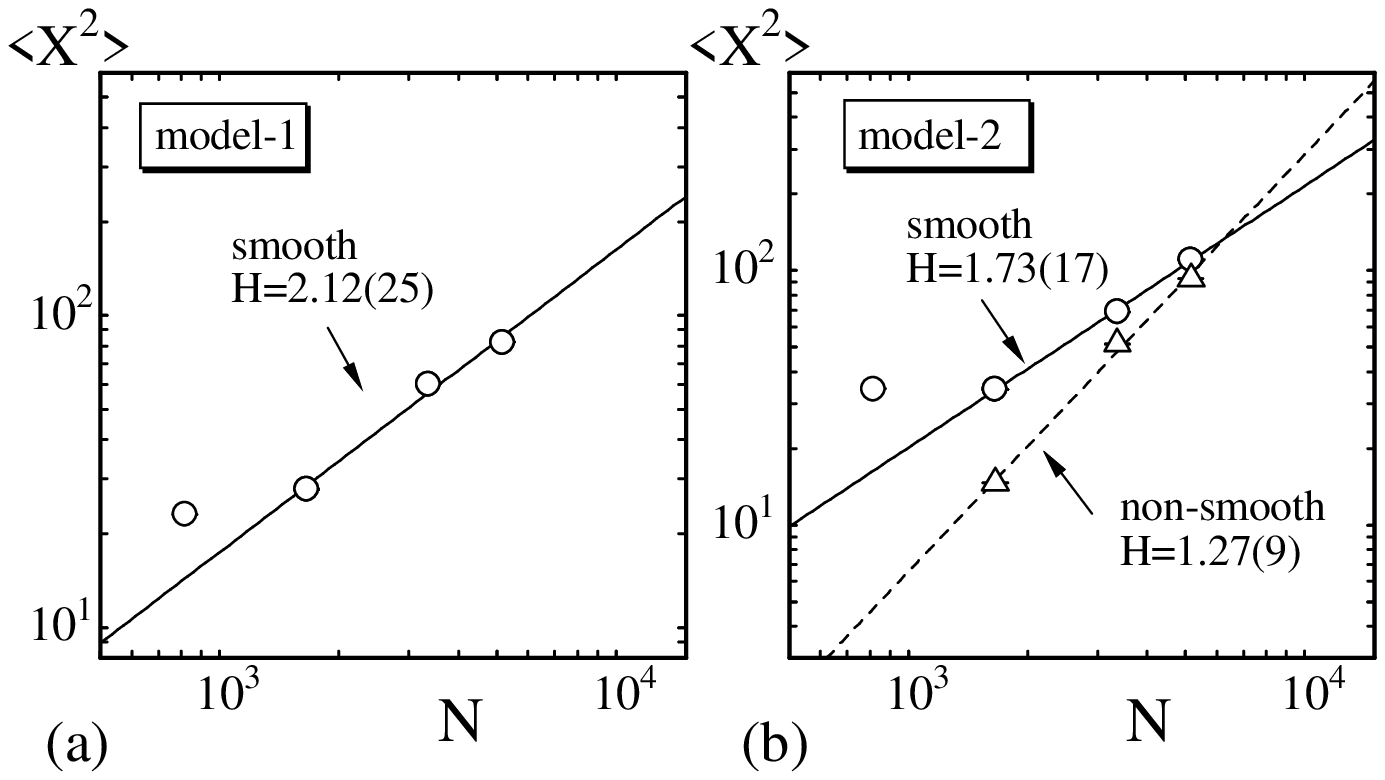}
}
\caption{(a) Log-log plots of $\langle X^2\rangle$ vs. $N$ ($\circ$) obtained at the smooth phase close to the transition point of model 1, (b) log-log plots of $\langle X^2\rangle$ vs. $N$ obtained in the smooth phase ($\circ$) and in the non-smooth phase ($\triangle$) close to the transition point of model 2. The straight lines and the dashed line are drawn by fitting the data to Eq. (\ref{Hausdorff}). }
\label{fig-7}
\end{figure}
%++++++++++++++++++++++++++++++++++
Figure \ref{fig-7}(a) shows log-log plots of $\langle X^2\rangle$ against $N$, where $\langle X^2\rangle$ are obtained by averaging $X^2$ over several points $\alpha$ close to the transition point with the weight of inverse error; $\alpha\!=\!24$, $\alpha\!=\!28$ for $N\!=\!5167$,  $\alpha\!=\!25$, $\alpha\!=\!30$, $\alpha\!=\!35$  for $N\!=\!3367$,  $\alpha\!=\!27$, $\alpha\!=\!28$, $\alpha\!=\!30$  for $N\!=\!1567$, and  $\alpha\!=\!37$, $\alpha\!=\!40$ for $N\!=\!817$. The reason why we need such averaging of $X^2$ is that $X^2$ changes with $\alpha$ in the smooth phase in contrast to the case of spherical surfaces \cite{KOIB-PRE-2005,KOIB-NPB-2006}. The straight line is obtained by fitting the largest three $X^2$  to
\begin{equation}
\label{Hausdorff}
X^2 \sim N^{2/H},
\end{equation}
where $H$ is the Hausdorff dimension. Thus we have 
\begin{equation}
\label{H-model-1}
H=2.12\pm 0.25, \; (\mbox{smooth})\qquad (\mbox{model 1}).
\end{equation}
The result $H\!=\!2.12(25)$ in Eq. (\ref{H-model-1}) is consistent with that the surface of model 1 is almost spherical in the smooth phase, as we have seen in Fig.\ref{fig-2}(c).

The Hausdorff dimension in the branched polymer phase is expected to be $H\!=\!2$, however, we can not extract information about $H$ in the non-smooth phase from our numerical results of model 1. One of the reason is because the non-smooth phase appears only on the surface of size $N\!=\!3367$ and $N\!=\!5167$, and another reason is because $X^2$ is not always constant against $\alpha$ in the non-smooth phase on those surfaces, as we can see in  Fig.\ref{fig-6}(a). If we could have performed very high-statistics simulations on larger surfaces, $H$ can be obtained even in the non-smooth phase.  

The solid line in Fig.\ref{fig-7}(b) is drawn by fitting $\langle X^2\rangle$ in the smooth phase of model 2; they were obtained by averaging $X^2$ close to the transition point; $\alpha\!=\!40$, $\alpha\!=\!44$ for $N\!=\!5167$,  $\alpha\!=\!40$, $\alpha\!=\!48$ for $N\!=\!3367$,  $\alpha\!=\!40$, $\alpha\!=\!44$, $\alpha\!=\!48$  for $N\!=\!1567$, and  $\alpha\!=\!55$, $\alpha\!=\!60$ for  $N\!=\!817$. Then, we have
\begin{equation}
\label{H-model-2}
H=1.73\pm 0.17,  \; (\mbox{smooth}).
\end{equation}
The result $H\!=\!1.73(17)$ is considered to represent a property characteristic to the smooth phase as in the case of model 1. In fact, $H\!=\!1.73(17)$ in (\ref{H-model-2}) is close to $H\!=\!2$, and therefore it is consistent with that the surface of model 2 is almost spherical in the smooth phase, as we have seen in Fig.\ref{fig-3}(d).  

The dashed line in Fig.\ref{fig-7}(b) is drawn by fitting $\langle X^2\rangle$ in the non-smooth phase of model 2; they were obtained by averaging $X^2$ close to the transition point; $\alpha\!=\!24$, $\alpha\!=\!28$, $\alpha\!=\!32$ for $N\!=\!5167$,  $\alpha\!=\!32$ for $N\!=\!3367$,  $\alpha\!=\!32$, $\alpha\!=\!36$  for $N\!=\!1567$. Then we have $H\!=\!1.27\pm0.09$, which implies that the non-smooth phase seems not to be the branched polymer phase at least in model 2. The non-smooth phase seems rather close to an oblong linear surface, because $H\!=\!1.27\pm0.09$ is close to $H\!=\!1$.

%++++++++++++++++++++++++++++++++++
\begin{figure}[hbt]
\centering
\resizebox{0.49\textwidth}{!}{%
  \includegraphics{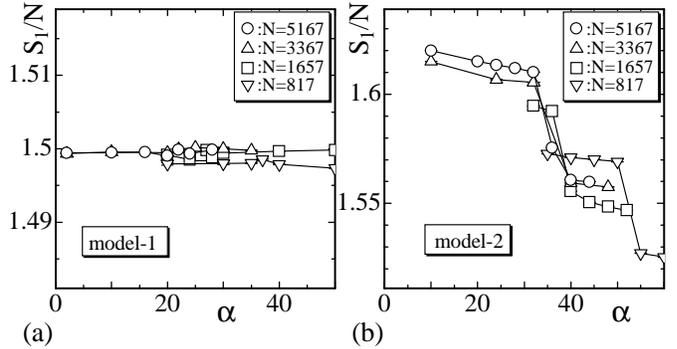}
}
\caption{(a) $S_1/N$ vs. $\alpha$ for model 1, and (b) $S_1/N$ vs. $\alpha$ for model 2. $S_1/N$ has a clear jump in (b).}
\label{fig-8}
\end{figure}
%++++++++++++++++++++++++++++++++++
Figures \ref{fig-8}(a) and \ref{fig-8}(b) show the Gaussian bond potential $S_1/N$ of model 1 and model 2, respectively, although $S_1$ is not included in the Hamiltonian of model 2. We find from Fig.\ref{fig-8}(a) that the relation $S_1/N\!=\!3/2$, which is expected from the scale invariant property of the partition function of model 1, is not influenced by the first-order transition.  Whereas in model 2 $S_1/N$ discontinuously changes at the transition point, where $S_2$ has a gap as shown in Fig.\ref{fig-4}(b). This discontinuity of $S_1$ also indicates that model 2 undergoes a discontinuous transition. Therefore, the discontinuous transition of model 2 is characterized by gaps of $S_2$ and $S_1$, although $X^2$ seems continuous at the transition point as we have seen in Fig.\ref{fig-6}(b).

%++++++++++++++++++++++++++++++++++
\begin{figure}[hbt]
\centering
\resizebox{0.49\textwidth}{!}{%
  \includegraphics{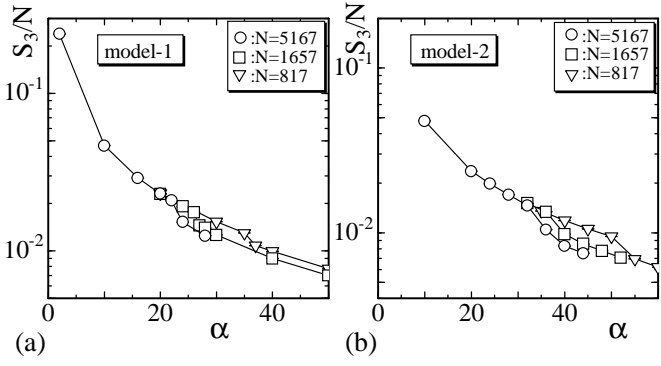} 
}
\caption{(a) $S_3/N$ vs. $\alpha$ for model 1, and (b) $S_3/N$ vs. $\alpha$ for model 2. }
\label{fig-9}
\end{figure}
%++++++++++++++++++++++++++++++++++
The intrinsic curvature energies $S_3/N$ of model 1 and model 2 are shown in Figs.\ref{fig-9}(a) and \ref{fig-9}(b) in a log-linear scale. The discontinuity of $S_3/N$ is relatively unclear compared with that of $S_2/N_B$ in Figs.\ref{fig-4}(a),\ref{fig-4}(b). Nevertheless, $S_3/N$ also changes discontinuously at the transition point in both models. Moreover, we find also that $S_3/N$ in the crumpled phase is quite larger than those in the other phases. These indicate that that both non-smooth surface and crumpled surface are non-flat as an intrisnsic curvature surface. Thus, the non-smooth surface and the crumpled surface are non-flat both extrinsically and intrinsically.

%++++++++++++++++++++++++++++++++++
\begin{figure}[hbt]
\centering
\resizebox{0.49\textwidth}{!}{%
  \includegraphics{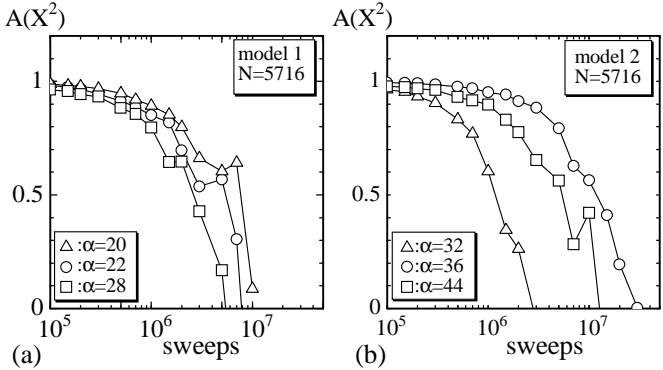}
}
\caption{Auto-correlation coefficient $A(X^2)$ of (a) model 1 at $\alpha\!=\!20$ (non-smooth), $\alpha\!=\!22$ (non-smooth), $\alpha\!=\!28$ (smooth phase), and (b) model 2 at  $\alpha\!=\!32$ (non-smooth), $\alpha\!=\!36$ (transient), $\alpha\!=\!44$ (smooth). 
}
\label{fig-10}
\end{figure}
%++++++++++++++++++++++++++++++++++
In order to see the convergence speed of the MC simulations in both models, we plot in Figs.\ref{fig-10}(a) and \ref{fig-10}(b) the autocorrelation coefficient $A(X^2)$ of $X^2$ defined by 
\begin{eqnarray}
A(X^2)= \frac{\sum_i X^2(\tau_{i}) X^2(\tau_{i+1})} 
   {  \left[\sum_i  X^2(\tau_i)\right]^2 },\\ \nonumber
 \tau_{i+1} = \tau_i + n \times 500, \quad n=1,2,\cdots. 
\end{eqnarray}
The horizontal axes in the figures represent $500\!\times\! n\;(n\!=\!1,2,\cdots)$-MCSs, which is a sampling-sweeps between the samples $X^2(\tau_i)$ and $X^2(\tau_{i+1})$. The coefficients $A(X^2)$ in Figs.\ref{fig-10}(a) and \ref{fig-10}(b) are obtained on the surface of size $N\!=\!5617$ in model 1 and model 2, respectively. 

 We clearly see from Fig.\ref{fig-10}(a) that the convergence speed in the smooth phase ($\alpha\!=\!28$) is almost identical to that in the non-smooth phase ($\alpha\!=\!20$), and it is also identical to that at $\alpha\!=\!22$, which seems very close to the transition point but remains in the non-smooth phase. The reason why the convergence speed in the smooth phase is comparable to the one in the non-smooth phase is that the phase space volume ($\subseteq {\bf R}^3$), where the vertices $X_i$ take their values, in the smooth phase is almost identical to that in the non-smooth phase. Figure \ref{fig-10}(b) shows the coefficient $A(X^2)$ of model 2. They were obtained in the non-smooth phase ($\alpha\!=\!32$), in the smooth phase ($\alpha\!=\!44$), and at a transient point ($\alpha\!=\!36$) between the two phases. We find that the convergence speed in the non-smooth phase ($\alpha\!=\!32$) is relatively faster than that in the smooth phase in model 2.   

Decorrelation MCS is about $0.3\!\times\! 10^7\sim 1\!\times\!10^7$ on the $N\!=\!5167$ surface close to the transition point in both models, because $A(X^2)\sim 0$ at $0.3\!\times\! 10^7\sim 1\!\times\!10^7$ in Figs.\ref{fig-10}(a),\ref{fig-10}(b). Therefore, the total number of MCS ($0.4\!\times\! 10^8\sim 1\!\times\!10^8$) after the thermalization, which was shown in Figs.\ref{fig-5}(a)--\ref{fig-5}(d), is considered to be sufficient, because it is very larger than the decorrelation MCS. However, we must note that the statistics is still not so high to obtain the Hausdorff dimension in the non-smooth phase and in the smooth phase as stated before. 

We should mention that phenomena of the critical slowing down in $A(X^2)$ at the transition point is very hard to see in our MC technique. The reason of this is because of very low speed of convergence. The change from the non-smooth phase to the smooth one appears to be almost irreversible. If the vertices are once trapped in a small region of ${\bf R}^3$ at the transition point, they hardly expand to be a smooth surface at least with the simulation technique in this paper. We obtain no series of MC data which includes configurations in the smooth phase and those in the non-smooth phase alternately even at the transition point. We consider that this irreversibility is only due to the simulation technique. These trapping phenomena were also seen in the same model on a sphere \cite{KOIB-EPJB-2004}, which was studied by using the same simulation technique. But, the transition between the two different phases must eventually be seen after long MC simulations at $\alpha$ sufficiently apart from the transition point.

Finally, we comment on the coordination number of surfaces. The maximum coordination number $q_{\rm max}$ is  $q_{\rm max}\!=\!22$ on the $N\!=\!5167$ surface of model 1, and $q_{\rm max}\!=\!19$ on the $N\!=\!5167$ surface of model 2. These values of $q_{\rm max}$ were almost independent of $\alpha$ at $2\leq\alpha\leq 28$ for model 1 and at $10\leq\alpha\leq 44$ for model 2. 

%------------------------------------------
\section{Summary and conclusion}\label{Conclusions}
%------------------------------------------
The purpose of this study was to understand how the fluidity of dynamical triangulation influences the first-order crumpling transition observed in the model on the fixed connectivity disk, which is an open surface. We were also interested in that the transition is independent of whether the surface is open or closed. 

Two types of fluid surface model have been investigated on a dynamically triangulated disk with intrinsic curvature; one with the Gaussian bond potential and the other with a hard-wall potential in place of the Gaussian bond potential. Both potentials are defined on all the bonds including the boundary bonds. The triangulated disk is obtained by dividing the edges of the hexagon into $L$-pieces, for a starting configuration of the MC simulations. The triangulated disk is characterized by $(N,N_E,N_T)\!=\!(3L^2\!+\!3L\!+\!1, 9L^2\!+\!3L, 6L^2)$, where $N_E,N_T$ are the total number of edges and the total number of triangles. In order to see whether the finite size effect influences the transition or not, we used surfaces of size up to $N\!=\!5762$, which is relatively larger than those previous studies on dynamically triangulated surfaces. The boundary bonds are prohibited from flipping, and the remaining internal bonds are allowed to flip so that at least one of the two-vertices of the bond connects to an internal vertex after the flip. 

We have shown that both models undergo a first-order phase transition between the smooth phase and the non-smooth phase. The transition in model 1 is characterized by a discontinuity of $S_2$ and that of $X^2$, whereas the transition in model 2 by a discontinuity of $S_2$ and that of $S_1$. Those discontinuities are considered to reflect a discontinuous transition. A phase transition is considered to be of first-order whenever some physical quantity has a gap or a jump. 

The transition occurs independent of whether the Gaussian bond potential is included in the Hamiltonian or not. Surfaces appear to be spherical in the smooth phase in both models. Our results indicate that the first-order transition is not influenced by the diffusion of vertices over dynamically triangulated surfaces if the intrinsic curvature is included in the Hamiltonian as the curvature term. Moreover, the results in this paper together with that in \cite{KOIB-EPJB-2004} indicate that the transition is independent of whether the surface is open or closed.  

It is interesting to study the model on the disk with grand canonical simulations so that the number of boundary vertices or bonds can change.

This work is supported in part by a Grant-in-Aid for Scientific Research, No. 15560160. The author (H.K.) acknowledges the staff of the Technical Support Center at Ibaraki National College of Technology for their help on computer analysis. 

%----------------------------------------------------------
%\vspace*{3mm}
%\noindent
%{\bf Acknowledgment}\\
%\vspace*{2mm}
%\par

%\vfill\eject

\end{document}